\documentclass[aps,prl,twocolumn,showpacs,preprintnumbers,amsmath,amssymb,superscriptaddress]{revtex4}

\usepackage{graphicx}
\usepackage{bm}
\begin{document}
\
\onecolumngrid 
 
\title{Search for $Z'\rightarrow e^{+}e^{-}$ Using Dielectron Mass and Angular Distribution}
\affiliation{Institute of Physics, Academia Sinica, Taipei, Taiwan 11529, Republic of China} 
\affiliation{Argonne National Laboratory, Argonne, Illinois 60439} 
\affiliation{Institut de Fisica d'Altes Energies, Universitat Autonoma de Barcelona, E-08193, Bellaterra (Barcelona), Spain} 
\affiliation{Baylor University, Waco, Texas  76798} 
\affiliation{Istituto Nazionale di Fisica Nucleare, University of Bologna, I-40127 Bologna, Italy} 
\affiliation{Brandeis University, Waltham, Massachusetts 02254} 
\affiliation{University of California, Davis, Davis, California  95616} 
\affiliation{University of California, Los Angeles, Los Angeles, California  90024} 
\affiliation{University of California, San Diego, La Jolla, California  92093} 
\affiliation{University of California, Santa Barbara, Santa Barbara, California 93106} 
\affiliation{Instituto de Fisica de Cantabria, CSIC-University of Cantabria, 39005 Santander, Spain} 
\affiliation{Carnegie Mellon University, Pittsburgh, PA  15213} 
\affiliation{Enrico Fermi Institute, University of Chicago, Chicago, Illinois 60637} 
\affiliation{Joint Institute for Nuclear Research, RU-141980 Dubna, Russia} 
\affiliation{Duke University, Durham, North Carolina  27708} 
\affiliation{Fermi National Accelerator Laboratory, Batavia, Illinois 60510} 
\affiliation{University of Florida, Gainesville, Florida  32611} 
\affiliation{Laboratori Nazionali di Frascati, Istituto Nazionale di Fisica Nucleare, I-00044 Frascati, Italy} 
\affiliation{University of Geneva, CH-1211 Geneva 4, Switzerland} 
\affiliation{Glasgow University, Glasgow G12 8QQ, United Kingdom} 
\affiliation{Harvard University, Cambridge, Massachusetts 02138} 
\affiliation{Division of High Energy Physics, Department of Physics, University of Helsinki and Helsinki Institute of Physics, FIN-00014, Helsinki, Finland} 
\affiliation{University of Illinois, Urbana, Illinois 61801} 
\affiliation{The Johns Hopkins University, Baltimore, Maryland 21218} 
\affiliation{Institut f\"{u}r Experimentelle Kernphysik, Universit\"{a}t Karlsruhe, 76128 Karlsruhe, Germany} 
\affiliation{High Energy Accelerator Research Organization (KEK), Tsukuba, Ibaraki 305, Japan} 
\affiliation{Center for High Energy Physics: Kyungpook National University, Taegu 702-701; Seoul National University, Seoul 151-742; and SungKyunKwan University, Suwon 440-746; Korea} 
\affiliation{Ernest Orlando Lawrence Berkeley National Laboratory, Berkeley, California 94720} 
\affiliation{University of Liverpool, Liverpool L69 7ZE, United Kingdom} 
\affiliation{University College London, London WC1E 6BT, United Kingdom} 
\affiliation{Centro de Investigaciones Energeticas Medioambientales y Tecnologicas, E-28040 Madrid, Spain} 
\affiliation{Massachusetts Institute of Technology, Cambridge, Massachusetts  02139} 
\affiliation{Institute of Particle Physics: McGill University, Montr\'{e}al, Canada H3A~2T8; and University of Toronto, Toronto, Canada M5S~1A7} 
\affiliation{University of Michigan, Ann Arbor, Michigan 48109} 
\affiliation{Michigan State University, East Lansing, Michigan  48824} 
\affiliation{Institution for Theoretical and Experimental Physics, ITEP, Moscow 117259, Russia} 
\affiliation{University of New Mexico, Albuquerque, New Mexico 87131} 
\affiliation{Northwestern University, Evanston, Illinois  60208} 
\affiliation{The Ohio State University, Columbus, Ohio  43210} 
\affiliation{Okayama University, Okayama 700-8530, Japan} 
\affiliation{Osaka City University, Osaka 588, Japan} 
\affiliation{University of Oxford, Oxford OX1 3RH, United Kingdom} 
\affiliation{University of Padova, Istituto Nazionale di Fisica Nucleare, Sezione di Padova-Trento, I-35131 Padova, Italy} 
\affiliation{LPNHE-Universite Pierre et Marie Curie-Paris 6, UMR7585, Paris F-75005 France; IN2P3-CNRS} 
\affiliation{University of Pennsylvania, Philadelphia, Pennsylvania 19104} 
\affiliation{Istituto Nazionale di Fisica Nucleare Pisa, Universities of Pisa, Siena and Scuola Normale Superiore, I-56127 Pisa, Italy} 
\affiliation{University of Pittsburgh, Pittsburgh, Pennsylvania 15260} 
\affiliation{Purdue University, West Lafayette, Indiana 47907} 
\affiliation{University of Rochester, Rochester, New York 14627} 
\affiliation{The Rockefeller University, New York, New York 10021} 
\affiliation{Istituto Nazionale di Fisica Nucleare, Sezione di Roma 1, University of Rome ``La Sapienza," I-00185 Roma, Italy} 
\affiliation{Rutgers University, Piscataway, New Jersey 08855} 
\affiliation{Texas A\&M University, College Station, Texas 77843} 
\affiliation{Istituto Nazionale di Fisica Nucleare, University of Trieste/\ Udine, Italy} 
\affiliation{University of Tsukuba, Tsukuba, Ibaraki 305, Japan} 
\affiliation{Tufts University, Medford, Massachusetts 02155} 
\affiliation{Waseda University, Tokyo 169, Japan} 
\affiliation{Wayne State University, Detroit, Michigan  48201} 
\affiliation{University of Wisconsin, Madison, Wisconsin 53706} 
\affiliation{Yale University, New Haven, Connecticut 06520} 
\author{A.~Abulencia}
\affiliation{University of Illinois, Urbana, Illinois 61801}
\author{D.~Acosta}
\affiliation{University of Florida, Gainesville, Florida  32611}
\author{J.~Adelman}
\affiliation{Enrico Fermi Institute, University of Chicago, Chicago, Illinois 60637}
\author{T.~Affolder}
\affiliation{University of California, Santa Barbara, Santa Barbara, California 93106}
\author{T.~Akimoto}
\affiliation{University of Tsukuba, Tsukuba, Ibaraki 305, Japan}
\author{M.G.~Albrow}
\affiliation{Fermi National Accelerator Laboratory, Batavia, Illinois 60510}
\author{D.~Ambrose}
\affiliation{Fermi National Accelerator Laboratory, Batavia, Illinois 60510}
\author{S.~Amerio}
\affiliation{University of Padova, Istituto Nazionale di Fisica Nucleare, Sezione di Padova-Trento, I-35131 Padova, Italy}
\author{D.~Amidei}
\affiliation{University of Michigan, Ann Arbor, Michigan 48109}
\author{A.~Anastassov}
\affiliation{Rutgers University, Piscataway, New Jersey 08855}
% ...
\collaboration{CDF Collaboration}
\noaffiliation

\begin{abstract}
We search for $Z^{\prime}$ bosons in dielectron events produced in 
$p\bar{p}$ collisions at $\sqrt{s}=1.96$ TeV, using a 0.45 fb$^{-1}$ 
dataset accumulated with the CDF II detector at the
Fermilab Tevatron. To identify the $Z^{\prime}\rightarrow e^{+}e^{-}$ signal, both the
dielectron invariant mass distribution and the angular distribution
of the electron pair are used. No evidence of a signal is found, and
95\% confidence level lower limits are set on the $Z^{\prime}$ mass for
several models. Limits are also placed on the mass and gauge coupling 
of a generic $Z^{\prime}$, as well as on the contact interaction mass scales
for different helicity structure scenarios.

\end{abstract}

\maketitle


Many extensions of the standard model (SM) gauge group  predict 
the existence of 
electrically-neutral, massive gauge bosons
commonly referred to as $Z^{\prime}$~\cite{rizzo, CDDT, heather, langaker, nima}.
The leptonic decays $Z'\rightarrow \ell^{+}\ell^{-}$ provide the most distinctive signature 
for observing the $Z^{\prime}$ signal at a hadron collider.
In two recent publications, the Collider Detector at Fermilab (CDF) Collaboration
has set limits on different $Z^{\prime}$ models by analyzing the invariant mass ($M_{\ell\ell}$)
spectrum of the dielectron, dimuon, and ditau final states, using a dataset corresponding to an
integrated luminosity of approximately 0.2 fb$^{-1}$~\cite{kaori, taus}. 
Besides the dilepton mass  $M_{\ell\ell}$, it has
been shown that the angular distribution of the dilepton events can also be used to
test the presence of a 
$Z^{\prime}$ boson by detecting its interference 
with the SM $Z$ boson~\cite{bodek}.
In this Letter, for the first time at a hadron collider, the
massive resonance search technique ($M_{\ell\ell}$ analysis) is extended to include dilepton angular 
information to detect $Z^{\prime}\rightarrow e^{+}e^{-}$ 
decays in 0.45 fb$^{-1}$ of data accumulated with the CDF II detector. 
As the integrated luminosity increases, 
the sensitivity of the standard $M_{\ell\ell}$ analysis plateaus;
adding the angular information starts to become an important handle for extending the 
$Z^{\prime}$ exclusion reach and discovery potential.
In this Letter many of the theoretical $Z^{\prime}$ models are surveyed, and
results are reported for the sequential $Z^{\prime}_{SM}$, the canonical superstring-inspired $E_{6}$
models $Z_{\chi}$, $Z_{\psi}$, $Z_{\eta}$, $Z_{I}$, $Z_{N}$, $Z_{\textrm{sec}}$~\cite{rosner,
paul2}, the ``littlest'' Higgs $Z_{H}$ model~\cite{nima, heather}, the four
generic  model-lines of Ref.~\cite{CDDT}, and contact-interaction
searches. No significant evidence of a $Z^{\prime}$ signal is found,
and the tightest constraints to date are set on most of these models.

The CDF detector is described in detail elsewhere~\cite{detector}. For this study, the 
relevant subdetector systems are the central tracking chamber (COT) and the central and the
plug calorimeters. The COT is a 96-layer open-cell drift chamber immersed in a 1.4 Tesla 
magnetic field, used to measure charged particle momenta within the pseudorapidity range
$|\eta|<1.0$~\cite{axes}. Surrounding the COT are the electromagnetic (EM) and hadronic (HAD) calorimeters,
segmented in projective $\eta-\phi$ towers pointing to the nominal collision point $z=0$. 
The central calorimeters measure the energies of particles within the range $|\eta|<1.1$, 
while the plug calorimeters extend the range to $1.1<|\eta|<3.6$. Two triggers were
used to select the data for this analysis. 
The main trigger requires two high-$E_{T}$ EM clusters in the calorimeter while a 
backup trigger accepts events with a single electron candidate with very high 
$E_{T}$ and looser electron-selection requirements.

Events are selected with two high-$E_{T}$ electron
candidates~\cite{wzc,greg}, of which at least one is required to have been measured in the central 
calorimeter.
A matching COT track is required for all central candidates. Events
with same-sign central electron pairs are rejected, and an isolation condition
for the energy found within a cone of angular radius 
$R=\sqrt{(\Delta\eta)^2+(\Delta\phi)^2}=0.4$ around the electron is imposed
for electron candidates. The angular distribution is measured using $\cos\theta$*,
where $\theta$* is the angle between the electron and the incoming quark in
the Collins-Soper frame~\cite{collins}. The search is performed using events
with $M_{ee}>200$~GeV/$c^2$.

$Z^{\prime}$ production is expected to interfere with the SM Drell-Yan $Z/\gamma^{*}$ 
process, altering the $\cos\theta$* distribution. For this reason, the $Z/\gamma^{*}$ 
process is not labeled as a ``background''. Instead, the $Z'/Z/\gamma$* is referred to as
the $Z^{\prime}$ $signal$, while the SM $Z/\gamma^{*}$ is referred to as the $SM$ $Drell$-$Yan$ production.
The term $background$ will be used to designate all other SM processes 
(excluding $Z /\gamma^{*}$) expected to contribute to the dielectron final 
state sample. Of these background sources, the most important are 
dijet events in which jets are misidentified as electrons, and diboson events 
(see Table \ref{t_results}). The dijet background is estimated using
the probability for a jet to be misidentified as an electron (``fake rate'')
which is measured in inclusive jet triggered data. The fake rate is then applied
to each jet in a sample of events containing one electron candidate and one or
more jets.
All non-dijet backgrounds are estimated using {\scriptsize{PYTHIA}}~\cite{pythia} 
Monte Carlo simulation, normalized to the product of the 
theoretical cross sections ~\cite{Baur93,Campbell} and the integrated luminosity. 
Other background processes such as $Z/\gamma$* 
$\rightarrow \tau^{+}\tau^{-}\rightarrow e^{+}e^{-} \nu_{\tau} \nu_{e} 
\bar{\nu_{\tau}} \bar{\nu_{e}}$ or pair production 
$t\bar{t}\rightarrow e^{+}e^{-} \nu_{e} \bar{\nu_{e}} b \bar{b}$ ~have negligible
contributions in the high-mass region considered for this analysis.

\begin{table}
\begin{center}
\begin{tabular}{|c|c|c|c|c|c|}
\hline
\hline
Source &$Z/\gamma$*$\rightarrow e^+ e^-$ & Dijet       & Diboson     &  Total
SM    & Observed   \\  
\hline
Events & $80.0\pm8.0$ &  $28_{-17}^{+14}$ & $6.8\pm1.4$ & $115_{-19}^{+16}$ & 120\\
 \hline
\hline
\end{tabular}
\caption{\label{t_results} Summary of the expected and 
observed numbers of dielectron events with $M_{ee}>200$ GeV/$c^2$ 
in 0.45 fb$^{-1}$.
This luminosity value was used to normalize the contributions from 
all processes with the exception of the dijet background.}
\end{center}
\end{table}

A leading-order calculation is used as the starting point to 
construct the signal and SM Drell-Yan Monte Carlo distributions~\cite{cata}. 
A next-to-next-to-leading order mass-dependent $K-$factor~\cite{CDDT}
is then included, 
followed by a ($M_{ee}$, $\cos\theta$*) parameterization of the CDF detector response~\cite{geant} 
to dielectron events.
This parameterization is extracted by running the full detector simulation on a 
large sample of $Z/\gamma^*\rightarrow e^+ e^-$  events generated with 
{\scriptsize{PYTHIA}} such that the distribution in $M_{ee}$ 
is roughly uniform between 0.03 and $1.05$ TeV/$c^2$.

For a particular $Z^{\prime}$ model (denoted by \texttt{\emph{H1}}), we isolate the $Z^{\prime}$ signal by 
using two variables:  the invariant mass of the electron-positron pair $M_{ee}$ in 
10~GeV/$c^2$ bins, and $\cos\theta$* in 0.25 bins.
The bidimensional distribution  $(M_{ee}, \cos\theta$*$)$ of the CDF data
is used to test two mutually exclusive hypotheses: 1) the $null$ hypothesis (\texttt{\emph{H0}}), 
where data are described by SM 
Drell-Yan and background distributions, and 2) the $test$ hypothesis 
(\texttt{\emph{H1}}), where data points are described by the $Z^{\prime}$ signal and background
distributions. A test statistic $Q$ is defined as:

\begin{align} \nonumber
Q(\overrightarrow{d}) & = -2\cdot \ln 
\frac{P(\overrightarrow{d}|\texttt{\emph{H1}})}{P(\overrightarrow{d}|\texttt{\emph{H0}})}
  = \mathrm{const}-2\cdot \sum_{i=1}^{N_{bin}}d_{i}\cdot 
 \ln\frac{{\cal N}_{i}^{\texttt{\emph{H1}}}}{{\cal N}_{i}^{\texttt{\emph{H0}}}}
\label{Qformula}
\end{align}
where $N_{bin}$ denotes the total number of bins in the $(M_{ee}, \cos\theta^{*})$ plane,
$\overrightarrow{d}=(d_{1},d_{2}, \dots, d_{N_{bin}})$ is the observed data distribution,
while ${\cal N}_{i}^{\texttt{\emph{H1}}}$ and ${\cal N}_{i}^{\texttt{\emph{H0}}}$ 
are the expected numbers of events in bin $i$ in the $\texttt{\emph{H1}}$ or 
$\texttt{\emph{H0}}$ hypotheses, respectively. 

Several sources of systematic uncertainty affect our measurement. First, a relative
uncertainty of 10\% on the total event rate is incurred primarily due to 
uncertainties in the luminosity measurement (6\%),
the dielectron detector acceptance and electron identification efficiency (5\%), 
and the LO calculation (4\%). 
The second dominating effect is the electron energy scale and resolution uncertainty, which modifies 
the shape of the
$M_{ee}$ and $\cos\theta$* distributions. The third source is the uncertainty in
the background (particularly dijet) estimations. The  
dijet prediction uncertainty is extracted  
from the differences in the fake rate measured in kinematically different jet samples.
Finally, the uncertainty related to the choice of the parton distribution functions set 
(CTEQ6M~\cite{cteq}) is evaluated using the Hessian method advocated in
Ref.~\cite{hess}, 
and found to have a negligible effect on our results. 

For each of the $Z^{\prime}$ models ($\texttt{\emph{H1}}$ test hypotheses) 
mentioned in the first paragraph, a large number of simulated experiments 
$\overrightarrow{d}$ are drawn either from the $\texttt{\emph{H1}}$ or 
the $\texttt{\emph{H0}}$ hypotheses, and the corresponding $Q(\overrightarrow{d})$ values are 
stored in two separate histograms $f^{\texttt{\emph{H1}}}(Q)$ and $f^{\texttt{\emph{H0}}}(Q)$, respectively. 
The systematic uncertainties are accounted
for as described in Ref.~\cite{clsref}. 
The $f^{\texttt{\emph{H1}}}(Q)$ and $f^{\texttt{\emph{H0}}}(Q)$ distributions are 
integrated in the region $Q > Q_{0}$, where $Q_{0}$ is the value measured 
using the $\overrightarrow{d}$ distribution of the CDF data. 
If the ratio of the two integrals is less than 5\%, then the 
$\texttt{\emph{H1}}$ model is excluded at 95\% confidence level (C.L.)~\cite{clsref}.

\begin{figure}[tb] \begin{center} 
\includegraphics[width=2.7in]{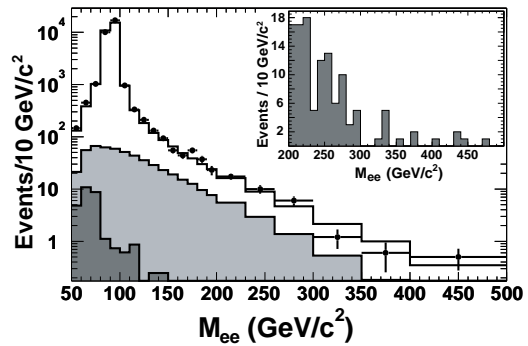}
\caption{\label{data1} $M_{ee}$ distribution of the data (points) compared 
to the prediction for SM Drell-Yan and backgrounds.
The individual contributions are stacked as follows: other backgrounds (dark grey),
dijet background (light grey), and SM Drell-Yan (open). 
The inset shows the $M_{ee}$ distribution of high-mass data events using a bin 
size of 10 GeV/$c^{2}$. There are no events in the sample with $M_{ee}>500$~GeV/$c^2$. }
\end{center} \end{figure}

\begin{figure}[tb] \begin{center} 
\includegraphics[width=2.6in]{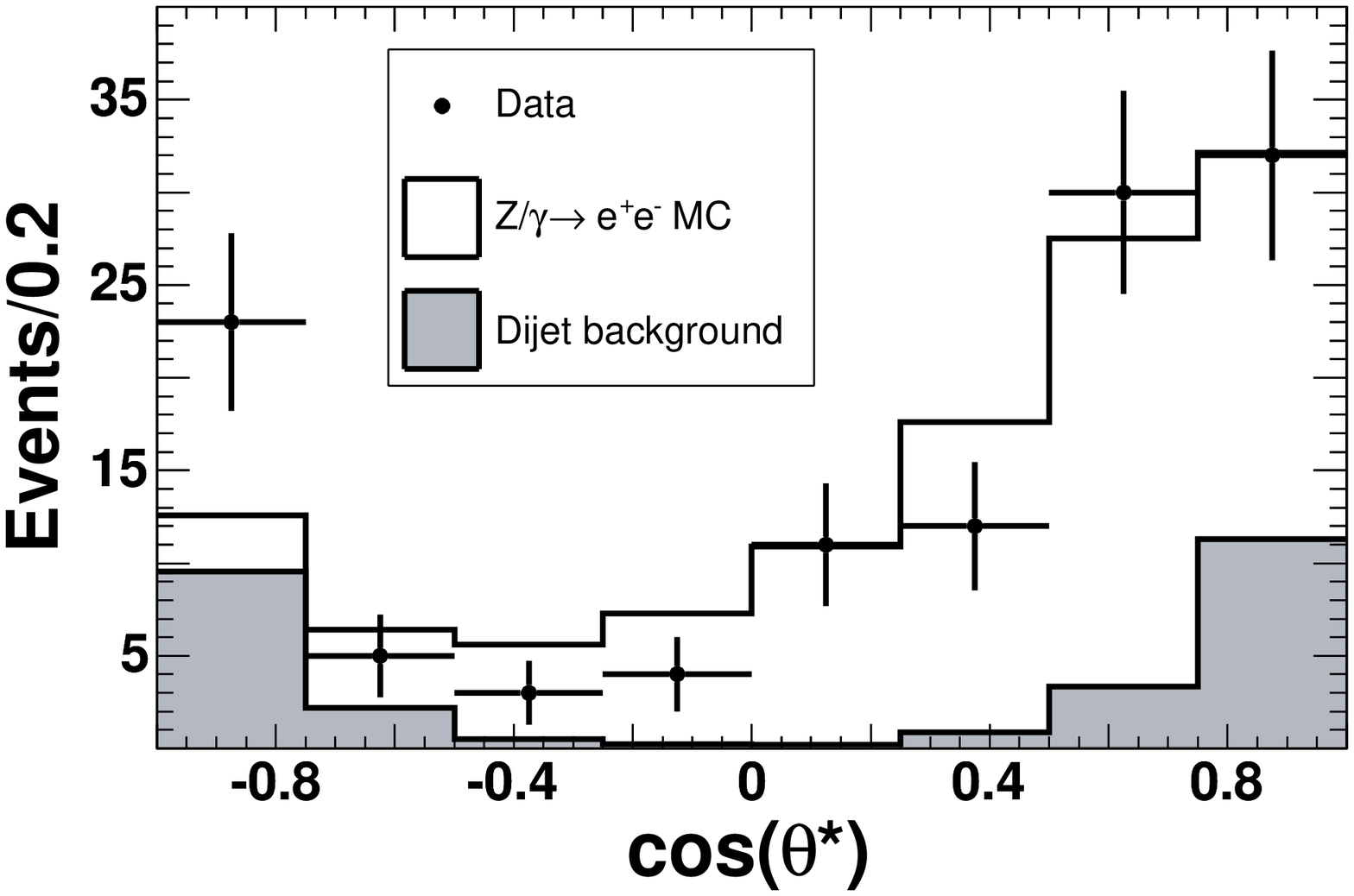}
\includegraphics[width=2.8in]{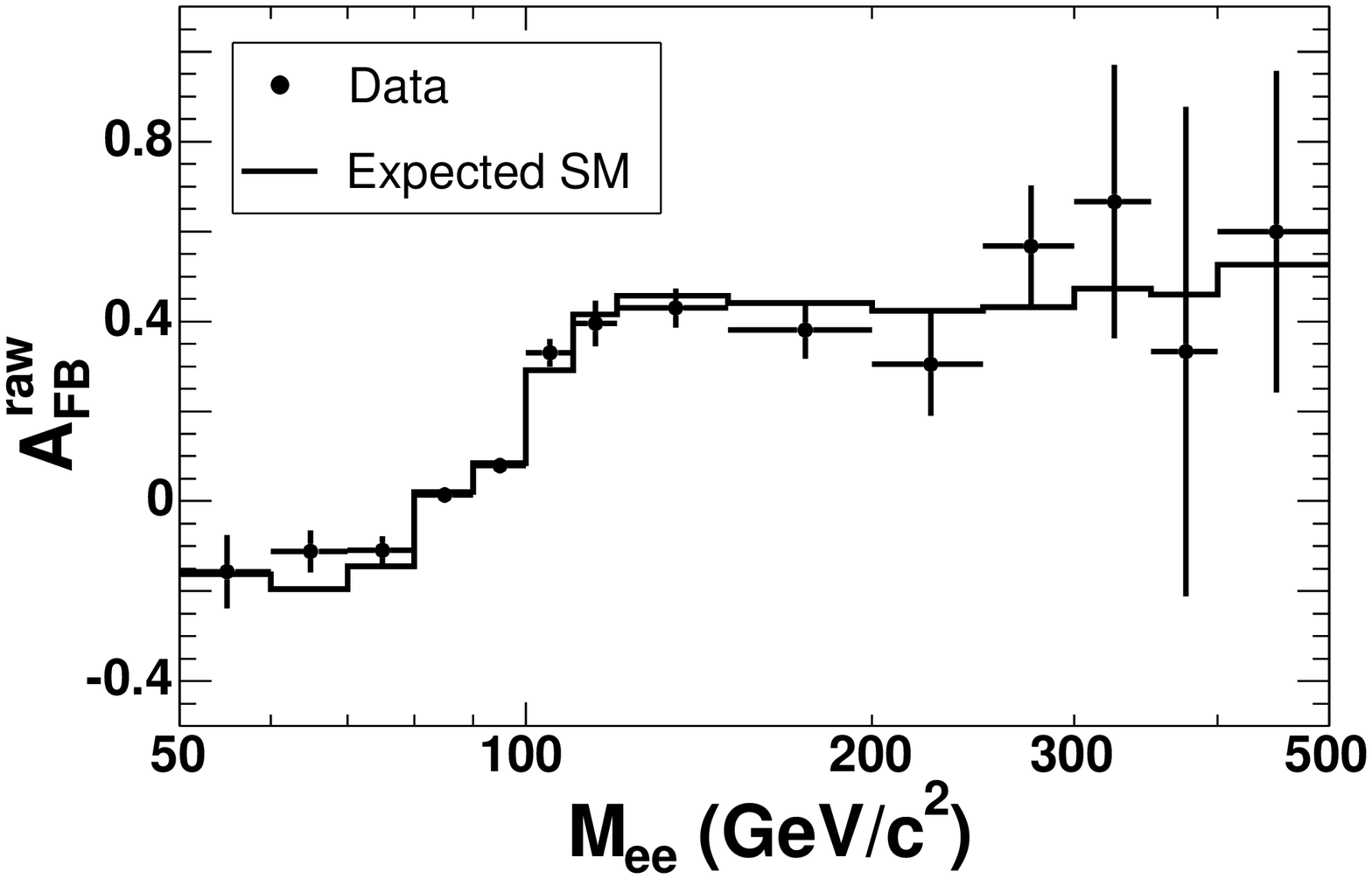}
\caption{\label{data2} 
Top: Distributions of $\cos\theta$* for the high mass region $M_{ee}>200$ GeV/$c^{2}$.
The points are the data, the open histograms are the predictions from 
Drell-Yan Monte Carlo simulation, and the shaded 
histograms are the background predictions.  The individual contributions
are stacked. Bottom: distributions of the 
forward-backward asymmetry $A^{raw}_{FB}$ for 
the data (points) and predicted SM processes (histogram).}
\end{center} \end{figure}

The CDF data is found to be consistent with the null (no $Z^{\prime}$) hypothesis, with a
probability $P(\overrightarrow{d}|\texttt{\emph{H0}})$ greater than the values obtained in 
87\% of $\texttt{\emph{H0}}$ simulated experiments. Figs. 
\ref{data1} and \ref{data2} show good agreement between data and Monte Carlo SM 
distributions for the $M_{ee}$ and $\cos\theta$* distributions. For illustration, Fig. 
\ref{data2} also presents the forward-backward
asymmetry $A^{raw}_{FB}$~\cite{greg} defined as $(N_{+}-N_{-})/(N_{+}+N_{-})$, where $N_{+}$ and $N_{-}$ 
are the numbers of forward ($\cos\theta$*$>0$) and backward 
($\cos\theta$*$<0$) events in the given $M_{ee}$ range. The superscript 
``raw'' is used here to emphasize that no detector acceptance, background subtraction,
or efficiency corrections are applied. The $A_{FB}$ plot
is a common way of representing the mass dependence of the angular distribution.

The sequential $Z^{\prime}_{SM}$ boson, which has the same couplings to fermions
as the SM $Z$ boson, is excluded by our data up to a mass of 850 GeV/$c^{2}$, at 
95\% C.L. It is noted 
here that using the dielectron invariant mass alone would require roughly 25\% more 
data for the same $Z^{\prime}_{SM}$ exclusion. In general, the improvement 
provided by including the $\cos\theta$* spectrum depends strongly on the particular $Z^{\prime}$
under investigation and the integrated luminosity being analyzed.
Other $Z^{\prime}$ theories include grand unification $E_{6}$ models~\cite{rosner,rizzo, langaker}, 
where the $E_{6}$ gauge group breaks down
as $E_{6}$$\rightarrow$SO(10)$\times$U(1)$_{\psi}$
$\rightarrow$SU(5)$\times$U(1)$_{\chi}\times$U(1)$_{\psi}$, and the SM gauge
structure results from breaking down the SU(5) group. Therefore,
an extra $Z^{\prime}$ will be a combination of the two U(1)'s: $Z_{\theta_{E_{6}}}=
Z_{\psi}\mathrm{sin}\theta_{E_{6}}+Z_{\chi}\cos\theta_{E_{6}}$. 
Table \ref{tab1} lists the 95\% C.L. 
lower limits on the $M_{Z^{\prime}}$ for the $Z_{\chi}$, $Z_{\psi}$,
$Z_{\eta}$, $Z_{I}$, $Z_{N}$, and the secluded $Z_{\textrm{sec}}$ $E_{6}$ models.

\begin{table*}[tb] 
\centering\small 
\begin{tabular}{|c|c|c|c|c|c|c|c|c|c|c|c|}
\hline\hline
 $Z^{\prime}$ Model	& $Z_{SM}$ &  $Z_{\chi}$ &  $Z_{\psi}$ &  $Z_{\eta}$ &  $Z_{I}$  &  $Z_{N}$ & $Z_{\textrm{sec}}$ &
 $Z^{0.3}_{H}$ & $Z^{0.5}_{H}$ & $Z^{0.7}_{H}$ & $Z^{1.0}_{H}$ \\
\hline
 Exp. limit (GeV/$c^{2}$)	& 860 &  735 & 725 & 745 & 650 & 710 & 675 & 625 & 765 & 835 & 910\\
\hline
 Obs. limit (GeV/$c^{2}$)	& 850 &  740 & 725 & 745 & 650 & 710 & 680 & 625 & 760 & 830 & 900\\
\hline\hline
\end{tabular} 
\caption{$Z^{\prime}$ exclusion summary: expected and observed 95\% C.L. lower limits on 
$M_{Z^{\prime}}$ for the sequential, the canonical $E_{6}$, and the littlest Higgs $Z^{\prime}$  models.
} 
\label{tab1} 
\end{table*}  

Another class of theories addressing electroweak symmetry breaking and the
hierarchy problem are the little Higgs theories~\cite{nima}, where 
$Z^{\prime}_{H}$ bosons are predicted in order to stabilize the Higgs mass against 
quadratically divergent one-loop corrections. In the minimal model of this type 
(the littlest Higgs), the $Z^{\prime}_{H}$ couples to left-handed fermions only, and these 
couplings are parameterized as functions of the mixing angle cotangent $\cot
\theta_{H}$~\cite{heather}.
Our results for $\cot \theta_{H} = 0.3$, 0.5, 0.7, and 1.0 are shown in Table \ref{tab1}, and
improve the results reported in~\cite{kaori}.

\begin{figure}[tb] \begin{center} 
\includegraphics[width=3.3in]{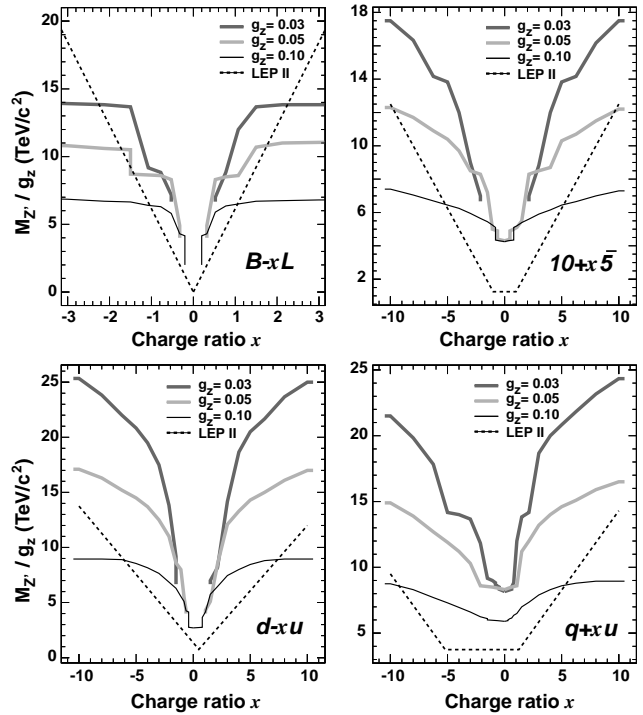}
\caption{\label{fig2} 
Exclusion contours for the $B-xL$, $10+x\bar{5}$, 
$d-xu$, and $q+xu$ $Z^{\prime}$ families. 
The dotted lines represent the
exclusion boundaries derived in Ref.~\cite{CDDT} from the LEP II results \cite{lep}. 
The region below each 
curve is excluded by our data at 95\% C.L.
Only models with $M_{Z'}>$200 GeV/$c^{2}$ are tested, which explains the gap at 
small $|x|$ for some models.}
\end{center} \end{figure}

A recent study reported in~\cite{CDDT} defines a more general
set of $Z^{\prime}$ models. This study uses simple 
constraints such as generation-independent fermion charges and gauge anomaly 
cancellations to reduce the number of parameters (17) required to define an 
arbitrary $Z^{\prime}$ model. 
Four families of models have been considered, each of them specified by three
parameters: the mass $M_{Z^{\prime}}$, the gauge coupling $g_{z}$, 
and the ratio $x$ of certain $U(1)$ charges. 
This three-dimensional parameter space is sampled to obtain the exclusion contours shown in 
Fig. \ref{fig2}. For small $|x|$ and $g_{z} \leq 0.10$, the exclusion limits are 
more stringent than the ones derived in Ref. \cite{CDDT} based on the LEP II results \cite{lep}. 

Finally, $Z^{\prime}$ mass constraints can be derived from searches for contact interactions, 
if the collider energy is far below the $Z^{\prime}$ pole~\cite{CDDT, lep, rochester}.
An effective Lagrangian for the $qqee$ contact interaction can be written as:
$\sum_{q} \sum_{i,j=L,R}
 4\pi\eta
 \bar{e_i}\gamma^{\mu}e_i \bar{q_j}\gamma_{\mu}q_j/{\Lambda^2_{ij}}$,
where $\Lambda$ is the scale of the interaction, and $\eta = \pm\ 1$
determines the interference structure with the $Z/\gamma^{*}$ amplitudes~\cite{ci}.
A generation-universal interaction is assumed and lower
limits are measured for $\Lambda$ in six helicity structure scenarios: LL, LR,
RL, RR, VV and AA (Table III)~\cite{vvva}.

\begin{table}[tb] 
\centering\small 
\begin{tabular}{|c|c|c|c|c|c|c|}
\hline\hline
 Interaction & LL & LR & RL & RR & VV & AA \\
\hline
 $\Lambda^{+}_{qe}$ limit (TeV/$c^2$) & 3.7 & 4.7 & 4.5 & 3.9 & 5.6 & 7.8 \\
\hline
 $\Lambda^{-}_{qe}$ limit (TeV/$c^2$) & 5.9 & 5.5 & 5.8 & 5.6 & 8.7 & 7.8 \\
\hline\hline
\end{tabular} 
\caption{95\% C.L. lower limits for the contact interaction mass scales.}
\label{contact} 
\end{table} 

In conclusion, we have searched for $Z^{\prime}$ decays to $e^{+}e^{-}$ pairs in
0.45 fb$^{-1}$ of data accumulated with the CDF II detector. To strengthen this search,
the reconstructed dielectron invariant mass $M_{ee}$ spectrum and the angular distribution 
of the electron pair $\cos\theta$*
are analyzed simultaneously. This is the first study of this kind at the Tevatron, and it
opens up a new avenue for exploring the $Z^{\prime}$ production in the inverse femtobarn 
luminosity regime. Many of the $Z^{\prime}$ models encountered in the literature are surveyed, no significant 
evidence for signal is found, and 95\% C.L. limits are set on these models.
Constraints are also placed on contact interaction mass scales far above the
Tevatron energy. Finally, exclusion contours for the generic $Z^{\prime}$ model-lines 
advocated in Ref.~\cite{CDDT} are mapped out. In comparison to the LEP contact interaction 
$Z^{\prime}$ search results given in~\cite{CDDT}, our results exhibit higher sensitivity 
in the small $|x|$ and small $g_{z}$ regions. 

We thank the Fermilab staff and the technical staffs of the participating 
institutions for their vital contributions. This work was supported by the 
U.S. Department of Energy and National Science Foundation; the Italian 
Istituto Nazionale di Fisica Nucleare; the Ministry of Education, 
Culture, Sports, Science and Technology of Japan; the Natural Sciences 
and Engineering Research Council of Canada; the National Science Council 
of the Republic of China; the Swiss National Science Foundation; the A.P.
Sloan Foundation; the Research Corporation; the Bundesministerium f\"ur Bildung und Forschung, Germany;
the Korean Science and Engineering Foundation and the Korean Research Foundation; 
the Particle Physics and Astronomy Research Council and the Royal Society, UK; the 
Russian Foundation for Basic Research; the Comisi\'on Interministerial de Ciencia 
y Tecnolog\'{\i}a, Spain; in part by the European Community's Human Potential 
Programme under contract HPRN-CT-2002-00292; and the Academy of Finland.
We thank M. Carena, B. Dobrescu, P. Langacker, H. Logan, and T. Tait for many 
fruitful discussions.


\begin{thebibliography}{99}
\bibitem{axes}In the CDF geometry, $\theta$ is the polar angle with respect to
the proton beam axis (positive $z$ direction), and $\phi$ is the azimuthal angle. The
pseudorapidity is $\eta=-\ln[\tan(\theta/2)]$. 
The transverse momentum, $p_{T}$, is the component of the momentum projected
onto the plane perpendicular to the beam axis. The transverse
energy $E_{T}$ of a shower or calorimeter tower is E $\sin\theta$, where E is 
the energy deposited.
\end{thebibliography}
\end{document}